\documentclass[]{aa}

\usepackage{amssymb,natbib,txfonts}

\bibstyle{aa}
\bibpunct{(}{)}{;}{a}{}{,}

\begin{document}

\voffset=-1cm
\def\be{\begin{equation}}
\def\ee{\end{equation}}
\def\bea{\begin{eqnarray}}
\def\eea{\end{eqnarray}}

\voffset=0.5cm

\title{A note on transition, turbulent length scales and transport in differentially rotating flows}

\author{ Denis Richard \and Sanford S. Davis}

\institute{ NASA Ames Research Center, MS 245-3, Moffett Field, California 94035}

\offprints{D.Richard / \email{drichard@mail.arc.nasa.gov}}

\authorrunning{Richard \& Davis}
\titlerunning{A note on differentially rotating flows}

\date{ Received ; Accepted}

\abstract{ In this note we address the issue of hydrodynamical instabilities in Astrophysical rotating shear flows in the light of recent publications focused on the possibility for differential rotation to trigger and sustain turbulence in the absence of a magnetic field. We wish to present in a synthetic form the major arguments in favor of this thesis along with a simple schematic scenario of the transition to and self-sustenance of such turbulence. We also propose that the turbulent diffusion length scale scales as the local Rossby number of the mean flow. A new prescription for the turbulent viscosity is introduced. This viscosity reduces to the so-called $\beta$-prescription in the case of velocity profiles with a constant Rossby number, which includes Keplerian rotating flows.
}

\maketitle

\section{Introduction}

If it is widely accepted that plane shear flows are subject to non-linear instability, the situation concerning their rotating counterpart is more controversial. Indeed, rotation alone is known to introduce constraints that could result in the stabilization of the flow. Differentially rotating flows are ubiquitous in astrophysics. From stellar interiors to accretion disks, from galactic rotation to planetary disks, they are met everywhere.While in many communities it is generally accepted that differential rotation could give rise to turbulence, the debate continues in the accretion disk community. Since the dawn of accretion disk modeling in the early 1970's, and in the absence of quantified transport properties supported by an identified physical mechanism, turbulence transport has been modeled with {\it ad hoc} prescriptions for the anomalous viscosity (the most popular being the so-called "alpha prescription"). Differential rotation has been proposed early as the possible source of this anomalous transport \citep{SS1973}. Recent numerical  simulations fail to show evidence of such instability in accretion disks. However, whether or not these simulations are adequate or sufficient to give a definitive answer to this issue remains open to question \citep{Longaretti,RichardThese,R2003}.
The difficulty of studying turbulence and instabilities has been reflected in the lack of a detailed physical mechanism along with dynamical and transport properties to apply in Astrophysical models. Several relevant instabilities are currently being investigated in the accretion disk community (\citet{Klahr}, \citet{Longaretti}, \citet{BH1991}), the most studied being the Magneto-Rotational Instability (MRI) \citep{BH1991}. This instability, linear in nature, is an excellent candidate as long as the disk is ionized enough, but is more unlikely to be efficient in cold disks such as proto-planetary nebulae (leading to the introduction of the "dead zone" layered accretion disk model by \citet{GammieDZ}).
The issue of differential rotation has been recently addressed with rather different approaches in several publication such as \citet{CZTL}, \citet{Longaretti}, \citet{LongarettiArxiv}, \citet{R2003} and \citet{TCZCL}.
It is our purpose to try to present these different results in a synthetic manner (section \ref{stab}) and to propose in their light a possible basic coherent scenario for the development of instabilities and the properties of the resulting turbulent state (section \ref{new}).




\section{Flow stability}
\label{stab}

 In this section we briefly comment on various published results concerning the stability of differentially rotating flows.

\subsection{The effect of rotation}

If plane shear flows are known to bear non-linear instabilities, the picture is quite different where the constraint of rotation is added. In a widely referred-to analysis, \citet{bhs} argued that hydrodynamic turbulence would be difficult to sustain in Keplerian flows without taking into account hydro-magnetic effects. Based on the equations of evolution of the turbulent velocity fluctuations energy, they deduced that rotating flows could not experience the turbulence due to the energy sink introduced by the Coriolis acceleration. Taking a closer look, one should note that the Coriolis force embodying the rotation influence on the fluid motion does not have any effect on the total energy equation. Its effect is to redistribute energy between the radial and azimuthal components of the motion, vanishing when these equations are summed. This is the natural consequence of the fictitious character of the  Coriolis force. Therefore, the rotation merely stiffens the system, add constraints,  but does not introduce any energy sink. The rotation and curvature of the flow introduces fundamental differences in comparison with plane shear flow. It does not necessary mean that the instability mechanisms present in plane flows disappear, but more likely that the stability and the characteristics of the resulting turbulence are modified (One can refer for example to \citet{LongarettiArxiv} for a comparative discussion on the role of rotation, curvature and viscous dissipation in plane and rotating shear flows).

\subsection{Finite amplitude perturbations and transition to turbulence}


In contrast with their linear counterpart, non-linear instabilities require the presence of finite (non-infinitesimal) amplitude perturbations to be triggered. \citet{CZTL} (and sequel \citet{TCZCL}) proposed the mechanism of transient growth and bypass mechanism (used for many years in the aerodynamics community)  to provide the required amplitudes.
Even in a linearly stable flow, the linear operator can provide transient amplification of perturbations. The results from \citet{CZTL} show that the initial perturbation amplitude can be greatly amplified. This amplification can introduce into the flow perturbations of amplitudes sufficient so that their non-linear interactions can no longer be neglected.

\citet{R2003} recently proposed a simple model describing the necessary conditions for self-sustained turbulence in differentially rotating flows, where it can be seen that the energy extraction is directly proportional to the shear present in the base flow (a classical result for shear flows). It also implies that there is a critical Rossby number ($Ro = {r \partial_r \Omega / 2 \Omega} $, where $r$ is the local radius and $\Omega$ is the mean rotation.) above which the energy extraction is sufficient to compensate for the stiffness introduced in the system by the mean rotation. It has to be pointed (as \citet{CZTL} also suggested) that the non-linear interactions (also referred to as turbulent diffusion) do not participate in the energy extraction, but only redistribute it and counteract the effects of rotation.  In this scenario, once the critical Rossby number has been reach (Richard 2003) and the critical amplitude is present, the flow can then undergo a transition from its laminar state to a state where non-linear shear turbulence is developed.



\section{Turbulent flow properties}
\label{new}

 \cite{R2003} focused on the stability properties of differentially rotating flows. Using a similar approach, we wish here to address the properties of the turbulent state of the bifurcated flow, in particular the turbulent transport and scale properties .

\subsection{Rossby number and turbulent scaling}

Rotation has a known tendency to constrain larger scales in a two-dimensional state while efficient turbulent diffusion is achieved by three-dimensional motions. Typical geophysical rotating flows exhibit both two-dimensional (at large scales) and three-dimensional (at smaller scales) structures. Within this picture, the relevant scale determining the turbulent diffusion should be the larger scale for which the turbulence can achieve to be three-dimensional.\citet{baroud2003} performed laboratory experiments on a rotating annulus, and have shown that low Rossby numbers are associated with two-dimensional turbulence whereas higher Rossby numbers (of order unity in their experiments) are associated with three-dimensional turbulent structures. The Rossby number for the mean flow can be re-written as

\be
Ro = \frac{1}{2} \cdot {r \over (\partial_r \ln \Omega)^{-1}},
\ee

where it appears as the ratio of local radius over the characteristic length scale of the shear. It can also be linked to a quantity more often referred to in astrophysics, the epicyclic frequency $\omega$, through the relation

\be
{\omega^2 \over 4 \Omega^2} = (1+Ro).
\ee

The Rossby number of a turbulent structure of characteristic length scale $\lambda$ and velocity $u$ ("turbulent Rossby number" hereafter), rotating with mean flow velocity, can be approximated by $Ro_{\lambda} \propto {r u / 2 \lambda^2 \Omega}$ : The denominator of the Rossby number is twice the rotation experienced by the turbulent structure, i.e. $2 (\Omega \pm u /r) \simeq 2\Omega$ ; The numerator is the derivative with respect to the radius of the turnover time, i.e. $r \partial_r \Omega + r \partial_r (u / \lambda) \simeq r \partial_r \Omega + r u / \lambda^2 \simeq r u / \lambda^2$, where we have used the relation

\be
{u / \lambda} \propto r \partial_r \Omega ,
\label{vorticity}
\ee

from \citet{R2003}, and the condition $\lambda << r$. We can then write, from Eq.(\ref{vorticity}) that

\be
Ro_{\lambda} \propto {r \over \lambda} Ro .
\label{Rolambda}
\ee

Hence, the ratio of the characteristic turbulent length scale over the local radius writes as

\be
{\lambda \over r} \propto {Ro \over Ro_{\lambda}} .
\label{lambdascaling}
\ee

From Eq. (\ref{Rolambda}), it follows that (for a given Rossby number associated with the mean flow) the turbulent Rossby number increases when going to smaller scales along the turbulent cascade. Remembering that high (resp. low) Rossby number are associated with three- (resp. two-) dimensional motions, this result is coherent with the classic picture of a turbulent spectrum showing two-dimensional structures at large scales and three-dimensional ones at smaller scales.

\subsection{Turbulent diffusion}

From an Astrophysical point of view, the main interest in studying turbulence might well be to characterize the turbulent transport of angular momentum or mass, by deriving the relevant model for the turbulent viscosity, which is classically written as

\be
\nu_t \propto u \cdot \lambda_{\nu} ,
\ee

where $\lambda_{\nu}$ stand for the largest turbulent scale participating in the diffusion process and $u$ its associated velocity. From Eqs. (\ref{vorticity}) and (\ref{lambdascaling}), it derives that

\be
\nu_t \propto \left( {Ro \over Ro_{\lambda_{\nu}}} \right)^{2} \cdot r^3 \partial_r \Omega .
\label{nut}
\ee

Two of the results from \citet{RZ99}, derived from experimental data from \citet{Wendt} and \citet{Taylor}, are :\\

- The ratio $\lambda_{\nu} / r$ is constant. From Eq.(\ref{lambdascaling}) it follows that $(Ro / Ro_{\lambda_{\nu}} )$ is a constant.\\

- The turbulent viscosity scales as $\nu_t = \beta \cdot r^3 \partial_r \Omega$, where $\beta$ is a numerical constant. This result combined with Eq.(\ref{nut}), implies that $\beta \propto  {(Ro / Ro_{\lambda_{\nu}}} )^{2}$.\\

Experiments from \citet{Taylor} used by \citet{RZ99} to derive the $\beta$ formulation of turbulent viscosity in rotating flows, all had the same maximum $Ro$. The reason is that the Rossby number within a laminar Couette-Taylor flow when the inner cylinder is at rest is given by $Ro = R_i / r$, where $R_i$ is the radius of the inner cylinder and r is the local radius. It follows that the absolute value of the Rossby number is maximum at the inner boundary of the flow and decreasing outward. In his classic experiments, Taylor modified the aspect ratio of his apparatus by changing the radius of the outer cylinder, hence did not modify the maximum Rossby number value in the flow. It follows that in Eq. (\ref{nut}), $Ro$ is to be treated as a constant for this set of experiments.
We conclude that both the apparent scaling of $\lambda_{\nu}$ and the constant value of the $\beta$ parameter derived in \citet{RZ99}, can be explained if we postulate that there exists a critical (transitional) $Ro_{\lambda_{\nu}}$ in the turbulent cascade above which the turbulence is three-dimensional and under which it is two-dimensional, and that its value is a constant that does not depend on the flow itself.
Concluding on the form of the turbulent viscosity, in the general case (from Eq.(\ref{nut})) it writes

\be
\nu_t \propto {1 \over {Ro_{\lambda_{\nu}}}} \cdot {(r^2 \partial_r \Omega)^3 \over 4 r \Omega^2}.
\ee

In the particular case of Keplerian flows, it reduces to

\be
\nu_t \propto \Omega_K \cdot r^2 ,
\ee

where $\Omega_K = \sqrt{GM / r^3}$ is the Keplerian angular velocity. This form is similar to the $\beta$-viscosity prescription, first introduced by \citet{Lynden} and more recently reintroduced by  \citet{RZ99} (based on laboratory fluid experiment data) and \citet{Duschl}, and applied by \citet{hrz}, \citet{Sandy}, \citet{Wehrstedt} and \citet{Lachaume}.

\section{Discussion}

Based on recent developments, we have proposed a simple scheme for the transition and self-sustenance of turbulence in non-magnetic differentially rotating flows. In this scenario, initial velocity fluctuations are amplified by their interaction with the mean shear (provided by the linear operator in the equations of motion). While this coupling between perturbations and averaged flow can not by itself destabilize the flow (due to its linearly stable nature) it can provide transient growth introducing finite amplitude fluctuations. These fluctuations in turn feed the non-linear interactions and trigger the instability, under the condition that the Rossby number of the mean flow has reached a critical value sufficient for the non-linear energy transfer to overcome the stiffness introduced by the mean rotation.
We conjecture that once the flow has become unstable, two- and three-dimensional turbulence coexist at different scales. The bi- or three-dimensional character of the motions is dictated by the value of the Rossby number at a given scale.
Finally we have introduced a new formulation for the turbulent viscosity of such turbulence, which reduces to the $\beta$ prescription in the case of constant Rossby number flows, including Keplerian rotation.

\begin{acknowledgements}
The authors wish to thank Dr. Jeffrey Cuzzi and Dr. Robert Hogan for their comments and support. D.Richard is supported by a Research Associateship from the National Research Council / National Academy of Sciences. This research has made use of NASA's Astrophysics Data System.
\end{acknowledgements}

\bibliographystyle{aa}
\bibliography{0519}

\end{document}